\documentstyle[11pt,newpasp,twoside,epsf]{article}
\def\gsim{\mathrel{\raise0.35ex\hbox{$\scriptstyle >$}\kern-0.6em
\lower0.40ex\hbox{{$\scriptstyle \sim$}}}}
\def\lsim{\mathrel{\raise0.35ex\hbox{$\scriptstyle <$}\kern-0.6em
\lower0.40ex\hbox{{$\scriptstyle \sim$}}}}

\reversemarginpar

\begin{document}
\title{History of mass assembly and star formation in clusters}
 \author{T.\ Kodama$^1$, Ian Smail$^2$,
 F.\ Nakata$^1$, S.\ Okamura$^1$, R.\ G.\ Bower$^2$
 }
\affil{$^1$Dept of Astronomy, Univ of Tokyo, 
Bunkyo-ku, Tokyo 113--0033, Japan\\
$^2$Dept of Physics, Univ of Durham, South Road,
Durham DH1 3LE, UK
}

\begin{abstract}
We present deep, panoramic multi-color imaging of the distant rich
cluster A\,851 ($z=0.41$) using Suprime-Cam on Subaru.
These images cover a 27$'$ field of view ($\sim 11 h_{50}^{-1}$\,Mpc),
and by exploiting photometric redshifts, we can isolate galaxies in a
narrow redshift slice at the cluster redshift. 
Using a sample of $\sim 2700$ probable cluster members ($<M_V^\ast+4$),
we trace the network of filaments and subclumps around the cluster core.
The depth of our observations, combined with the identification of filamentary
structure, gives us an unprecedented opportunity to test the influence
of the environment on the properties of low luminosity galaxies.  We
find an abrupt change in the colors of faint galaxies ($>M_V^\ast+1$) at
a local density of 100 gal.\,Mpc$^{-2}$.
The transition in the color-local density behavior
occurs at densities corresponding to subclumps within the filaments
surrounding the cluster.  Identifying the sites where the transition
occurs brings us much closer to understanding the mechanisms which are
responsible for establishing the present-day relationship between
environment and galaxy characteristics.
\end{abstract}

\section{Introduction}

Clusters of galaxies are continuously growing through the accretion of
galaxies and groups from the field.
The star formation activity in the accreting galaxies must be quenched
during the assimilation of the galaxies into the cluster.
This transformation is a key process in creating the
environmental dependence of galaxy properties, and may also underpin the
observed evolution of galaxy properties in distant clusters [e.g.\ 1.2]
However, the physical mechanism which is responsible for these changes
has not yet been identified [3,4,5].

The advent of Suprime-Cam, a revolutionary wide-field camera on
the Subaru telescope, has opened a new window in this field.
Its 27$'$ field of view can trace the variation of galaxy properties
from the cluster cores out to the surrounding field in an attempt to
identify the environment where the decline in the star formation in
accreted galaxies begins.

\section{Large scale structure}

As a first step towards a systematic study of distant clusters with Subaru
and Suprime-Cam,
we obtained deep ($\sim M^*+4$) $BVRI$ imaging of the rich cluster A\,851
at z=0.4.
We constructed an I-band selected sample which contains 15,055
galaxies brighter than $I=23.4$.

In order to assign cluster membership across our large field and to
faint magnitudes,
we applied photometric redshift technique [6]
to isolate the galaxies in a narrow redshift slice around the cluster redshift.
By applying the redshift cut of $0.32<z_{\rm phot}<0.48$, we can reduce the
field contamination by factor 10 at $I=23.4$, while keeping most of the
cluster members (80\%, from the estimate using the spectroscopic members [7]).

After this photometric redshift selection, it is straightforward
to map out the distribution of galaxies which are likely to be associated
with the cluster.
As shown in Fig. 1, several large scale structures are visible around
the cluster core, comprising many clumps and filamentary extensions
coming out directly from the core.
Importantly, most of these extensions from the core are aligned with
the surrounding subclumps.

\begin{figure}
\plotfiddle{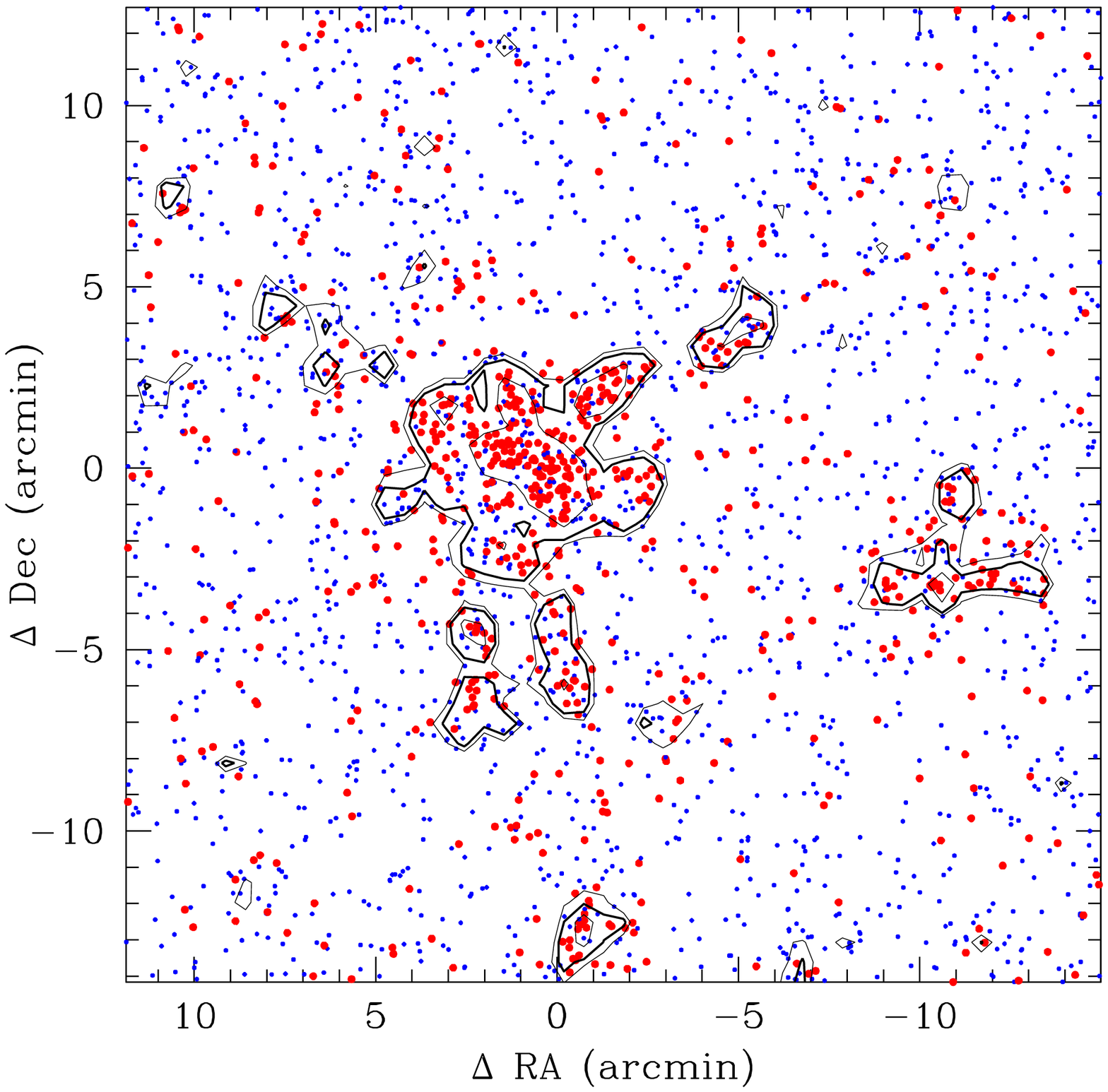}{6.5cm}{0.0}{0.35}{0.35}{-100}{-50.0}
\caption{
A wide-field map of A\,851 from our Suprime-Cam $BV\! RI$ imaging.
The field is 11\,Mpc on a side.  The points identify cluster members based
on photometric redshifts.  The remaining field contamination
is not subtracted.  The large and small circles separate the
colors of galaxies at the color 0.5 mag bluer than the color-magnitude
sequence in $B-I$, where large ones are in red.
The contours indicate the local surface density of
the member galaxies, corresponding to
$\log_{10}\Sigma=1.8$, 2.0 (thick) and 2.5\,Mpc$^{-2}$ (after correcting
for field contamination), respectively.  The thick contour
traces the boundary where the color distribution of galaxies changes
dramatically.
Large scale structure is clearly visible.
The overall shape resembles an `octopus', with a round head and many legs.
}
\end{figure}

The structures identified in this region are qualitatively similar to
those seen in cosmological simulations of the growth of clusters which
exhibit the filamentary/clumpy substructures on similar scales [e.g.\ 8]
It appears therefore that we are witnessing
A\,851 as it assembles through the accretion of galaxies and groups
along the filaments onto the cluster core from the surrounding field.

\subsection{Environmental dependence of galaxy properties}

By exploiting the striking large scale structure around this cluster, we
can investigate the influence of environment on the photometric properties
of galaxies.
We define the environment for each galaxy using the local surface number
density, $\Sigma$, of members, calculated from the 10 nearest neighbors.
We statistically subtract residual field contamination 
using a Subaru blank field data [9].

We divide the density distribution into three regimes as indicated
in Fig.~2a.  There is a close correspondence between
the local density and structure: the high density region corresponds
to the cluster core within $\lsim 1$\,Mpc; the medium density region
includes the structures defining the filaments surrounding the cluster;
and the low density region comprising the rest of the volume (Fig.~1).

\begin{figure}
\plottwo{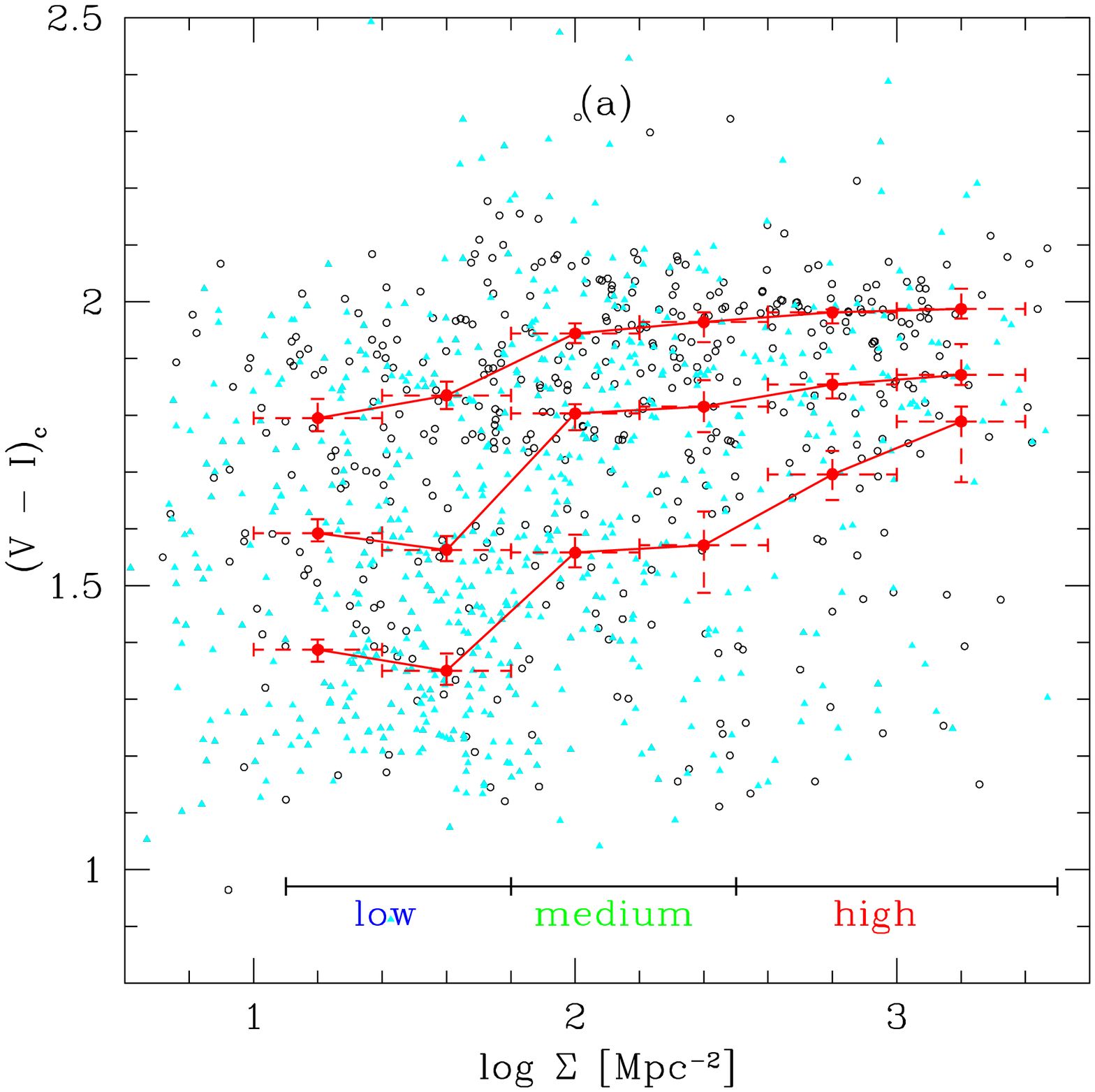}{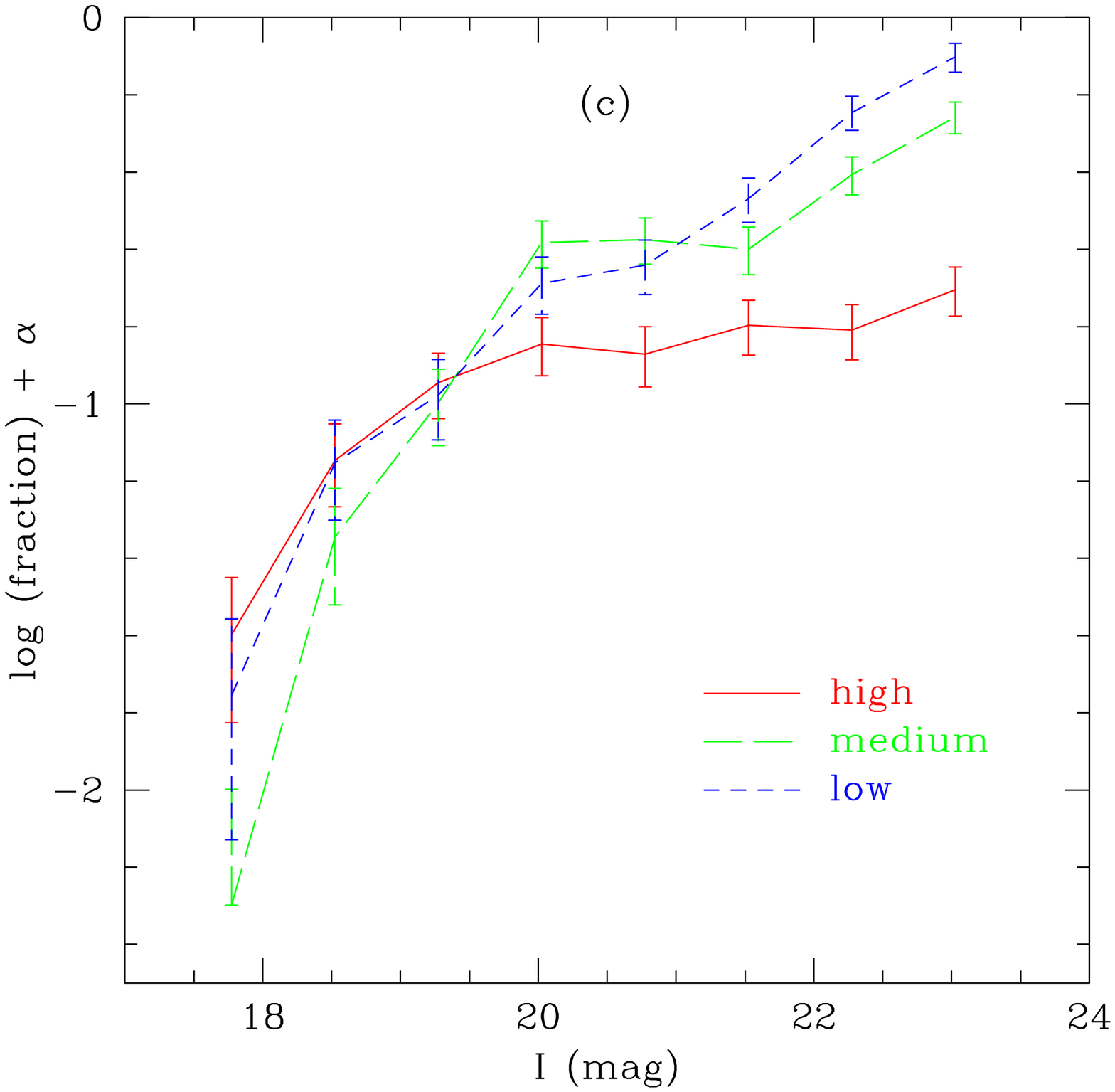}
\caption{
(a) The variation in color versus local galaxy density, $\Sigma$,
for cluster members brighter than $I=23.4$.  The open circles and
filled triangles show the galaxies brighter or fainter than $I=21.4$
($M_V^{\ast}$+2), respectively.  The three red lines represent the
loci of the 25, 50, and 75th percentile colors.
The colors are corrected for the slope of the color-magnitude relation.
We also correct for the residual field contamination using the blank
field data.
(c) The field-corrected luminosity function for cluster
galaxies in the three density slices from Panel (a).  The distributions
are normalized at $I=19.4$ ($M_V^\ast$) to match the high density curve.
A clear steepening of the luminosity function is visible at lower
densities.
}
\end{figure}

As shown in Fig.~2a, the color distribution in the high density region
is strongly peaked at $(V-I)_c\sim 2$, the color of an early-type cluster
member.  However, as we move to lower densities, the distribution becomes
dramatically bluer.
Notably, this color transition with local density occurs
quite abruptly at $\log_{10}\Sigma\sim 2$, indicating a threshold effect
in transforming galaxy properties.
The boundary corresponding to this critical density is highlighted in Fig.~1.
Surprisingly, this transition occurs in subclumps well outside the core.

Our ability to pin-point this environment is a fundamental step towards
identifying the dominant mechanism behind the environmental dependence
of galaxy properties.
Since the ram-pressure stripping is largely supressed in groups [4],
the effective mechanism at this hierarchy is either
galaxy--galaxy collisions (which cause cold disk gas to
be driven to the galaxy center creating a star burst [3]);
and `suffocation' (where {\it warm} gas in the galaxy's halo
is shock heated by the intra-cluster medium so that it can no longer
cool and replenish the cold gas in the disk [10]).
Our data do not distinguish directly between these possibilities,
but they can be distinguished by tracing the variation in the individual
components of galaxies, bulges and disks, between the cluster and
the field.

It is important to note that the changes in galaxy properties as a
function of local density are most prominent in faint galaxies ($>M_V^\ast+1$).
This is mirrored to the dramatic change
in the shape of the luminosity function with local galaxy density: where
we see a steepening of the faint end slope of the luminosity function
with decreasing density (Fig.~2c).
Assuming that their star formation effectively ceases, the relative
absence of low-luminosity galaxies at projected densities of
$\log_{10}\Sigma> 2$ can be understood as they will fade by $\gsim 1$\,mag
as their star formation declines [2].
This will put many of their descendents below our magnitude limit.
This is also qualitatively consistent with the
presence of large numbers of very low luminosity, $\gg M_V^\ast+3$, passive
dwarf galaxies in local clusters [11]

\acknowledgments

We are grateful to Dr Y.\ Komiyama for his assistance during our
observations.  We acknowledge the Suprime-Cam team
for allowing us to use the blank field data.  We also thank Drs.\ M.\
Balogh, N.\ Arimoto and K.\ Shimasaku for helpful discussion.  TK and FN
acknowledge the Japan Society for the Promotion of Science for support
through its Research Fellowships for Young Scientists.  IRS acknowledges
support from the Royal Society and the Leverhulme Trust.


\begin{references}
\reference [1] Butcher, H., Oemler, A., 1984, ApJ, 285, 426
\reference [2] Kodama, T., Bower, R. G., 2001, MNRAS, 321, 18
\reference [3] Moore, B., Katz, N., Lake, G., Dressler, A., Oemler, A., 1996,
    Nature, 379, 613
\reference [4] Adabi, M. G., Bower, R. G., Navarro, J. F., 2000, MNRAS,
    314, 759
\reference [5] Balogh, M. L., Navarro, J. F., Morris, S. L., 2000, ApJ,
    540, 113
\reference [6] Kodama, T., Bell, E. F., Bower, R. G., 1999, MNRAS, 302, 152
\reference [7] Dressler, A., Smail, I., Poggianti, B. M., Butcher, H.,
    Couch, W. J., Ellis, R. S.,  Oemler, A., 1999, ApJS, 122, 51
\reference [8] Ghigna, S., Moore, B., Governato, F., Lake, G., Quinn, T.,
    Stadel, J., 1998, MNRAS, 300, 146
\reference [9] Ouchi, M., et al., 2001, ApJ, 558, L83
\reference [10] Larson, R. B., Tinsley, B. M.,  Caldwell, C. N., 1980, ApJ, 237, 692
\reference [11] Binggeli, B., Sandage, A., Tammann, G. A., 1988, ARA\&A, 26, 509
\end{references}
\end{document}